\documentstyle[11pt,aaspp4,flushrt]{article} 

\begin{document}

\def\spose#1{\hbox to 0pt{#1\hss}}
\def\lta{\mathrel{\spose{\lower 3pt\hbox{$\mathchar"218$}}
     \raise 2.0pt\hbox{$\mathchar"13C$}}}
\def\gta{\mathrel{\spose{\lower 3pt\hbox{$\mathchar"218$}}
     \raise 2.0pt\hbox{$\mathchar"13E$}}}
\def\Msun{{\rm M}_\odot}
\def\msun{{\rm M}_\odot}
\def\Rsun{{\rm R}_\odot}
\def\Lsun{{\rm L}_\odot}
\def\half{{1\over2}}
\def\RL{R_{\rm L}}
\def\zs{\zeta_{s}}
\def\zR{\zeta_{\rm R}}
\def\dJJ{{\dot J\over J}}
\def\dMM{{\dot M_2\over M_2}}
\def\tKH{t_{\rm KH}}
\def\eck#1{\left\lbrack #1 \right\rbrack}
\def\rund#1{\left( #1 \right)}
\def\wave#1{\left\lbrace #1 \right\rbrace}
\def\dd{{\rm d}}
%
%
\def\DS{\displaystyle}
\def\se{\displaystyle{s_{\rm eff}}}
\def\tdr{\DS{t_{\rm dr}}}
\def\tce{\DS{t_{\rm ce}}}
\def\rt1{\DS{r_t-1}}
\def\dzsR{\DS{\zeta_s-\zR}}

\title{MASS TRANSFER CYCLES IN CLOSE BINARIES WITH EVOLVED COMPANIONS}
\author{A. R. King, \altaffilmark{1}, J. Frank, \altaffilmark{2}
U. Kolb \altaffilmark{1} and H. Ritter \altaffilmark{3}}

\altaffiltext{1} {Astronomy Group, University of Leicester, 
Leicester LE1 7RH, U.K. (ark@star.le.ac.uk, uck@star.le.ac.uk)}

\altaffiltext{2} {Department of Physics and Astronomy, Louisiana State
University,   Baton Rouge, LA 70803-4001, USA
       (frank@rouge.phys.lsu.edu)}

\altaffiltext{3} {Max--Planck--Institut f\"ur Astrophysik, 
Karl--Schwarzschild--Str. 1, D~85748 Garching, Germany
(hsr@MPA-Garching.MPG.DE)}

\begin{abstract}

We give a global analysis of mass transfer variations in 
low-mass X-ray binaries and cataclysmic variables 
whose evolution is driven by the nuclear expansion of the secondary star.
We show that limit cycles caused by 
irradiation of the secondary by the accreting primary are possible in
a large class of these binaries. In the high state the companion 
transfers a large fraction of its envelope mass on a thermal timescale.
In most cases this implies super--Eddington transfer rates, and would
thus probably lead to common--envelope evolution and the formation of 
an ultrashort--period binary. Observed systems with (sub)giant 
secondaries stabilize themselves against this possibility either by 
being transient, or by shielding the secondary from irradiation in some
way.

\end{abstract}


\section{INTRODUCTION}

Semidetached binaries in which a compact object (white dwarf, neutron star
or black hole)
accretes material via Roche lobe overflow from a companion 
on or near the main sequence are of great interest in current 
astrophysics. The evolution of such systems
is driven by orbital angular momentum losses via
gravitational radiation and magnetic braking 
(see e.g. King 1988 for
a review). Thus most properties of the binary, particularly the mean
mass transfer rate, depend essentially only on the secondary mass $M_2$
and change on the timescale 
$t_J \sim 10^8 - 10^9$ yr for angular momentum loss.
In a recent paper (King, Frank, Kolb \& Ritter 1996a, henceforth Paper I) 
we discussed the conditions under which mass
transfer can vary cyclically about the evolutionary mean
in these systems by developing 
a general formalism allowing one to study the stability of mass
transfer in systems driven by angular momentum losses. The existence of
cycles is required to account for the wide dispersion of e.g. mass transfer
rates at a given orbital period.
We concluded
that the most likely cause of such cycles is weak irradiation of the 
companion star by the accreting component. The conditions for this appear
to be fulfilled in a large class of cataclysmic variables (CVs), in
which the accreting star is a white dwarf.

Of course semidetached evolution is not restricted to systems with 
main--sequence companions. In systems with orbital periods $P\gta 1$ day
the orbital evolution and
hence the rate of mass transfer is either determined or strongly
influenced by nuclear evolution of the companion.
In this paper we discuss the stability of mass transfer in
systems containing a giant or a subgiant companion and consider the
possibility of irradiation--driven mass transfer cycles similar to
those thought to exist in CVs. For this purpose we generalize the
analysis of Paper I to include the effects of nuclear evolution on the
radius variations by using a simple core--envelope model for the companion.
This is a good representation of systems with low--mass giant
secondaries, which constitute the great majority of long--period compact 
binaries. However this description
does not apply to the recently--discovered
black--hole transient GRO J1655-40, where the companion star appears to
be crossing the Hertzsprung gap (Orosz \& Bailyn, 1996).

\section{GLOBAL ANALYSIS OF MASS TRANSFER VARIATIONS}

In this section we follow closely the formalism developed in Paper I,
casting the equations governing time--dependent mass transfer
in a semidetached binary in a form suitable for global analysis 
including the effects of nuclear evolution. We restrict
our analysis to the effect of
variations in the radius $R_2$ of the lobe--filling star on the transfer
rate, as this is the simplest way of causing mass transfer variations
(other ways can easily be accomodated: Paper I gives an example).
Note that this part of the analysis is quite general, in that there is
no presumption at this point 
that the radius variations result from irradiation.

Radius changes can
result from dynamical, thermal or secular processes. For example,
local adjustments in the structure of the atmospheric layers can take place
on the shortest (dynamical) timescale, while adjustments 
of the stellar radius
in response to secular mass loss and nuclear evolution occur on the longest 
(secular) driving timescale $t_{\rm dr}$ defined more precisely below. 
We represent the radius variation as
\begin{equation}
{\dot R_2 \over R_2} = \zeta_s \dMM + K_{\rm th}(R_2, \dot M_2)
                                    + K_{\rm nuc}\; .
\label{e2.1}
\end{equation}
Here
$\zeta_s$ is the adiabatic mass--radius exponent ($\simeq -1/3$ for a
fully convective star), $K_{\rm th}(R_2, \dot M_2)$ represents radius
variations due to thermal relaxation and irradiation of the star by the
primary, and $K_{\rm nuc}=(\partial\ln R_2/\partial t)_{\rm nuc}$ represents
the secular changes due to nuclear evolution.
We considered examples of specific forms of $K(R_2, \dot M_2)$ 
in Paper I and will discuss these again later.
The change of the mass transfer rate is given by
\begin{equation}
\ddot M_2 = {\dot M_2\over H} (\dot R_2 - \dot\RL) \approx
\dot M_2 {R_2\over H}\biggl({\dot R_2\over R_2}
- {\dot\RL\over\RL}\biggl)\; ,
\label{e2.2}
\end{equation}

\noindent
where $\RL$ is the critical Roche radius, and $H$ is the pressure
scale height in the secondary star's atmosphere. The
approximation given by
(\ref{e2.2}) is justified
since even for giant companions $\vert R_2 - \RL\vert
<< \RL, R_2$. The response of the Roche radius to mass loss is
described by

\begin{equation}
   {\dot\RL\over\RL} =  \zR\dMM + 2\dJJ\; ,
\label{e2.3}
\end{equation}

\noindent
where $\zR$ is a function of the mass ratio $M_2/M_1$, approximately
given by $\zR\approx 2 M_2/M_1 - 5/3$ for conservative mass transfer,
and $\dot J$ is the rate of loss
of orbital angular momentum.
Inserting (\ref{e2.1}) and (\ref{e2.3}) into (\ref{e2.2}), we obtain

\begin{equation}
\ddot M_2 = \dot M_2 {R_2\over H}
\biggl[ (\zeta_s-\zR)\dMM + K_{\rm th} + K_{\rm nuc} - 2\dJJ \biggl]\; .
\label{e2.4}
\end{equation}

For the more general case considered here the evolution of the
binary is driven by the combined effects of nuclear evolution and
angular momentum losses. We introduce the effective driving
timescale $t_{\rm dr}$, as follows,

\begin{equation}
{1\over t_{\rm dr}} =  {1\over t_{\rm nuc}} + {2\over t_J}\; ,
\label{e2.5}
\end{equation}

\noindent
where $t_{\rm nuc} = K_{\rm nuc}^{-1}$ is the nuclear timescale, and 
$t_J = -J/\dot J$ is the timescale for angular momentum losses.
The nuclear timescale is a strong function of the core mass, and is
$\sim 10^7 - 10^9$ yr for giants and much longer for low mass
main sequence dwarfs. The angular momentum loss time scale $t_J$ is
typically $\sim 10^8 - 10^9$ yr for main sequence  companions, 
but could be much longer for evolved
companions. Thus orbital evolution is driven mainly by angular momentum
losses in systems with main sequence companions (e.g. CVs) and nuclear
processes in systems with evolved companions. The introduction of
$t_{\rm dr}$ allows us to treat both cases simultaneously.
We define a dimensionless mass transfer rate

\begin{equation}
x = {-\dot M_2\over(-\dot M_2)_{\rm ad}}= -\dMM (\zeta_s-\zR) t_{\rm dr}\; ,
\label{e2.6}
\end{equation}

\noindent
where $(-\dot M_2)_{\rm ad}$ is the adiabatic mass transfer rate i.e.
the steady rate implied by equation (\ref{e2.4}) with $K_{\rm th}=0$.  
A system undergoing stable mass
transfer will typically do so at $x\sim 1$ (see Sect. 3).
We also introduce the dimensionless stellar radius

\begin{equation}
r =  R_2/R_e\; ,
\label{e2.7}
\end{equation}

\noindent
where $R_e = R_e(M_2,M_c)$ is the radius of the secondary in thermal 
equilibrium for a given total mass $M_2$ and a core mass $M_c$.
The radii and luminosities of lower giant--branch stars are virtually
independent of the mass of the envelope but depend strongly on the
mass of the degenerate helium core (Refsdal \& Weigert 1970). We allow
$R_e$ formally to depend on the total mass because then the equations
obtained are identical in form to those derived in 
Paper I and have wider applicability. 

As we discuss in Sect. 3, the secular equilibrium radius
of the companion under stable mass transfer differs slightly from $R_e$.
Although the equilibrium radius $R_e$ is only attained in the 
absence of mass transfer, we write formally

\begin{equation}
{\dot R_e\over R_e} = \zeta_e {\dot M_2\over M_2} + K_{\rm nuc}\; ,
\label{e2.8}
\end{equation}

\noindent
where $\zeta_e$ is defined by the above equation.
Taking the time derivative of equation (\ref{e2.6}), 
and neglecting secular variations of $H$, 
we obtain

\begin{equation}
-\ddot M_2 =  {M_2\over(\zs-\zR)t_{\rm dr} } \dot x + 
{\dot M_2 x\over (\zs-\zR)t_{\rm dr}}
+ \dot M_2 {{\rm d}\ln{[(\zs-\zR)t_{\rm dr} ]}\over {\rm d}t}\; .
\label{e2.9}
\end{equation}

Finally, we introduce $\epsilon = H/R_e$, which is
typically $\simeq 10^{-3}$ for lower giant branch stars (see below). We 
define
the dimensionless thermal relaxation function 
$p(r,x) = K_{\rm th}t_{\rm dr}$, and
the dimensionless time $\tau = t/(\epsilon t_{\rm dr})$.
With these assumptions and definitions (\ref{e2.9}) becomes

\begin{equation}
{{\rm d}x\over {\rm d}\tau} =  rx [1 + p(r,x) -x]
+ \epsilon x\biggl[{x\over (\zs-\zR)} 
+ t_{\rm dr} {\dd\ln{[(\zs-\zR)t_{\rm dr}]}\over\dd t}\biggr]\; .
\label{e2.10}
\end{equation}
\noindent
Note that since the terms inside the second set of
square brackets are of order unity and 
are multiplied by $\epsilon$, they can be safely neglected. Thus 
(\ref{e2.10}) reduces to

\begin{equation}
{{\rm d}x\over {\rm d}\tau} =  rx [1 + p(r,x) -x]\; .
\label{e2.11}
\end{equation}
With the variations of the equilibrium radius given by equation
(\ref{e2.8}), and using the above conventions, the radius equation 
(\ref{e2.1}) becomes

\begin{equation}
{{\rm d}r\over {\rm d}\tau} =
\epsilon r \biggl[ p(r,x) - {x\over\lambda} \biggr]\; ,
\label{e2.12}
\end{equation}

\noindent
where $\lambda = (\zeta_s - \zR)/(\zeta_s-\zeta_e)$. The
equations (\ref{e2.11}, \ref{e2.12}) describing radius and mass
transfer variations are identical in form to those derived in Paper I.
Thus we can take over from Paper I all the results for the
phase plane motion of the system, the critical curves
$\dot r = 0$, $\dot x = 0$, the fixed point(s) $(r_0,x_0)$ 
at the intersection(s) of these critical curves, the
stability anaylsis for the fixed point(s), and the conditions for limit
cycles, with the modification that now $t_{\rm dr}$ is a more
general driving time which includes both systemic angular momentum
losses and nuclear evolution. In the limit $t_{\rm nuc}\rightarrow\infty$
we recover the case studied in Paper I. 

From equations (\ref{e2.11}, \ref{e2.12}) it is easy to see that the 
stationary (secular mean) mass transfer rate at the fixed point
$x_0=(\zeta_s-\zR)/(\zeta_e-\zR)$ depends on the properties of
the companion and the driving rate. 
In particular it is independent of $p(r,x)$ (as it must be); 
and in most realistic cases $x_0\sim 1$. The radius $r_0$ at the fixed
point on the other hand is given implicitly in terms of $p(r,x)$; again
we typically have $r_0\sim 1$ (see Paper I for further details).

\section{BINARIES WITH SUBGIANT AND GIANT COMPANIONS}

 In their study of the evolution of compact binaries
containing a lower giant branch companion, Webbink, Rappaport and
Savonije (1983) introduced simple analytic expressions for the
luminosity and radius in terms of the core mass. For our purposes
it suffices to take only the first two terms of their
approximate formulae for the radius

\begin{equation}
R_e(M_c) = 12.55\Rsun \biggl({M_c\over 0.25\Msun}\biggr)^{5.1}\; ,
\label{e3.1}
\end{equation}
and the luminosity
\begin{equation}
L_e(M_c) = 33.1\Lsun \biggl({M_c\over 0.25\Msun}\biggr)^{8.1}\; ,
\label{e3.2}
\end{equation}
in thermal equilibrium. Using these equations
we can readily estimate $\epsilon$ for these stars
\begin{equation}
\epsilon={H\over R_e}= 2.1\times 10^{-3}\mu^{-1}
\biggl({M_c\over 0.25\Msun}\biggr)^{4.6} \biggl({M_2\over\Msun}
\biggr)^{-1}\; ,
\label{e3.2.1}
\end{equation}
where $\mu$ is the mean molecular weight in the stellar atmosphere.
From (\ref{e3.1}, \ref{e3.2}) the nuclear timescale for radial variations
is $t_{\rm nuc}= M_c/(5.1 \dot M_c)$, where $\dot M_c$ is calculated from 
the equilibrium luminosity assuming a hydrogen mass fraction of 0.7 and an
energy yield of $6\times 10^{18}$ erg g$^{-1}$. Thus 
\begin{equation}
t_{\rm nuc} = 1.0\times 10^8 {\rm yr} 
\biggl({M_c\over 0.25\Msun}\biggr)^{-7.1} \; ,
\label{e3.2.2}
\end{equation}
The Kelvin-Helmholtz timescale is
defined for the whole star as $\tKH=GM_2^2/(R_eL_e)$ so
\begin{equation}
\tKH = 7.6\times 10^4 {\rm yr}
\biggl({M_2\over \Msun}\biggr)^2
\biggl({M_c\over 0.25\Msun}\biggr)^{-13.2} \; .
\label{e3.2.3}
\end{equation}
The ratio of these timescales plays a major role in
the discussion of thermal relaxation and stability. We define
\begin{equation}
\rho_{\rm nuc} = {t_{\rm nuc}\over\tKH} = 1.4\times 10^3
\biggl({M_2\over \Msun}\biggr)^{-2} 
\biggl({M_c\over 0.25\Msun}\biggr)^{6.1} \; .
\label{e3.2.4}
\end{equation}

Since no explicit dependence of $R_e$ 
on the total mass of the companion $M_2$ is allowed in equations
(\ref{e3.1}, \ref{e3.2}),
it appears that $\zeta_e =0$. However, both computed models of
lower giant branch stars and analytic treatment of nearly fully 
convective stars following the approximations used by Kippenhahn and 
Weigert (1990) for the Hayashi line show a weak dependence on the
total stellar mass $M_2$. Taking a photospheric opacity of the form 
$\kappa = \kappa_0 P^a T^b$,
and adopting a polytropic index $n=3/2$ for the envelope,
one can show that 
\begin{equation}
\zeta_e = -{a+3\over 5.5a+b+1.5}\; .
\label{e3.3}
\end{equation}
With $a=1$, $b=3$, the above expression yields $\zeta_e=-0.4$, while
direct estimates from numerical stellar models indicate $\zeta_e
\approx -0.2$ to $-0.3$. We adopt the latter range in describing the 
reaction of the star to mass loss, although the relation (\ref{e3.1})
is adequate for computing the nuclear and thermal timescales.

We now describe the effects of thermal relaxation and irradiation 
on the radius of the evolved companion. We do this here
using a simple homologous model and making the same approximations
as in King et al. 1995 and in 
Paper I with slight changes appropriate for giants:

1) Homology relations can be used to describe the stellar structure, i.e.\
the effective temperature $T_*$ on the unirradiated portion of the 
stellar surface remains essentially constant (Hayashi line) and 
any stellar radius change changes the
surface luminosity as $L \propto R_2^2$. In constrast
$L_{\rm nuc}$ remains unchanged, as given by equation (\ref{e3.2}).

2) If a fraction $s$ of
the stellar surface is exposed to a uniform irradiating flux $F$, the
luminosity of the star, i.e.\ the loss of energy per unit time from
the interior, is reduced by the blocking luminosity
\begin{equation}
L_b = s L \; \tanh \left\{ k \; {F\over F_*} \right\} \; , 
\label{e3.4}
\end{equation}
where $F_* = \sigma T^4_*$ is the unperturbed stellar flux and $k$ a 
parameter which is adjusted to 
approximate the results of more realistic model calculations 
[e.g.\ those by Ritter et al.\ (1996a, 1996b)]. 
The above ansatz for $L_b$ is motivated by the facts that a) $L_b = 0$
if $F = 0$, b) $L_b \to s L$ if $F\gg F_*$, and c) the transition between
$F=0$ and $F\gg F_*$ is smooth and monotonic. With these assumptions we 
get the thermal relaxation function including irradiation effects 
(cf.\ equation (8) of King et al.\ 1995 and equation (38) of Paper I)
\begin{equation}
p(r,x) = - f \rho \; \left\{ \left[ 1- s\, \tanh 
\rund{{x\over x_c}} \right] \, r^3 - r \right\} \; . 
\label{e3.5}
\end{equation}
Note that the quantity $\rho = t_{\rm dr}/\tKH$ introduced above differs 
from $\rho=t_J/\tKH$ used in Paper~I [see equation (\ref{e2.5})].
The ratio $f = \tKH/t_{\rm ce}$, with $t_{\rm ce}$ being the thermal timescale
for the convective envelope, is a constant  
which depends on the internal structure (i.e. mainly on the
ratio of envelope mass to the total mass of 
the secondary), and
\begin{equation}
x_c = 
{2(\zeta_s-\zR) \rho \over k\eta} \; {M_2 \over M_1} \; {R_1 \over R_e} \; 
\rund{{a\over R_e}}^2.
\label{e3.6}
\end{equation}
This is the dimensionless critical mass transfer rate which yields an
irradiating flux $F=F_*/k$. Cycles typically occur provided that $x_c$
is $\lta$ the secular mean transfer rate $x_0$. In (\ref{e3.6}) 
$a$ is the orbital separation and $\eta$ a dimensionless
efficiency factor relating the irradiating flux $F$ to the mass transfer
rate via
\begin{equation}
F = {\eta \over 8\pi} \; {GM_1 (-\dot M_2) \over R_1 a^2} \; . 
\label{e3.7}
\end{equation}

The critical mass transfer rate for CVs is
$x_c \sim 0.1\rho/k\eta\sim 1-10$, where the value of $k$ 
depends on the type of companion star assumed, while $\eta$ depends
mainly on the radiation spectrum
(typical values are $0.1 \lta \rho \lta 5, 0.2 \lta
k \lta 0.9$, with $\eta \sim $ a few percent). As this critical rate 
is of order the secular mean 
mass transfer rate $x_0 \sim 1$, irradiation can
potentially drive cycles in CVs, as we found in paper I. 
For LMXBs with similar
mass ratios, the much smaller neutron star or effective black--hole radius 
implies $x_c\sim 0.0001\rho/k\eta\sim 0.1-1$. 
Thus we can potentially expect
cycles in these systems also. This contrasts with the case of 
LMXBs with unevolved companions (cf Paper I), where the companion is
too strongly irradiated to give cycles (critical transfer rates
$x_c << x_0 \sim 1$). 
For LMXBs with evolved companions the large ratio $\rho$ of
driving to thermal timescales almost compensates the effect of the
smaller primary radius to give $x_c \sim 0.1 - 1$.

\section{STABILITY OF MASS TRANSFER FROM AN EVOLVED COMPANION}

The simple thermal relaxation function given in the 
previous section allows us to discuss the stability of mass 
transfer in close binaries with evolved companions. More detailed 
calculations using the bipolytrope approximation (Ritter, Zhang \&
Kolb 1996b) or integrating
the full equations of stellar structure with appropriate changes
in the boundary conditions (Hameury \& Ritter 1996) produce results
which to a first approximation can be represented by this 
analytic form with suitable $k, x_c \sim 1$.
In Paper I we noted that a necessary condition for 
limit cycles is that the fixed point is unstable,   
which in turn requires $p_r<0$ and $p_x>1$ simultaneously. 
These derivatives are easily calculable for
the simple form of equation 
(\ref{e3.5}), namely
\begin{equation}
p_r(r_0,x_0) = - f \rho \; \left\{ \left[ 1- s\, \tanh 
\rund{{x_0\over x_c}} \right] \,3 r_0^2 - 1 \right\} \; , 
\label{e4.1}
\end{equation}
and
\begin{equation}
p_x(r_0,x_0) = {f \rho s\over x_c} r_0^3 {\rm sech}^2
{\rund{{x_0\over x_c}}}\; . 
\label{e4.2}
\end{equation}
For physically reasonable situations $s<0.5$ and $r_0>1$, and thus 
clearly $p_r<0$ as required. The second condition $p_x > 1$
for instability can be rewritten as in Paper I:
\begin{equation}
s {x_0\over x_c}{\rm sech}^2{\rund{{x_0\over x_c}}} > 
{x_0\over f\rho r_0^3}\; ,
\label{e4.3}
\end{equation}
where the l.h.s. is a function which has a maximum value of 
$\approx 0.448 s$ at $x_0/x_c\approx 0.78$ and vanishes at both small
and large $x_0/x_c$. As emphasised in Paper I the r.h.s. of equation 
(\ref{e4.3}) depends
mainly on the type of companion and the driving mechanism. Clearly the
large values of $\rho$ for giant or subgiant companions makes these systems
extremely vulnerable to the irradiation instability. Note
that if $s=0$ in (\ref{e4.3}), i.e. irradiation is not included, 
or the flux is somehow blocked (e.g.
by a thick disk), the inequality is violated and mass transfer
driven by nuclear evolution is stable as assumed in conventional 
treatments of these systems.

Thus the simple model described above suggests that most
binaries with irradiated 
lower giant branch--companions cannot transfer mass
at a stable rate. More detailed models discussed later confirm this result.
While the linearized equations are adequate to discuss the stability
of the fixed point, as in Paper I, we need the full 
non-linear evolution
equations to understand the ultimate fate of the system. It turns out
that the phase point representing the evolution
of the system away from an unstable fixed point is trapped in its 
vicinity, and will
settle into a limit cycle which is described in more detail below.

\section{PROPERTIES OF THE LIMIT CYCLE}

The simple
thermal relaxation functions $p(r,x)$ given by (\ref{e3.5})
for evolved companions and  
equation (38) of Paper I for main sequence companions yield 
approximate analytic expressions for a number of properties of
the limit cycle. The two cases can be treated simultaneously
by taking 
\begin{equation}
p(r,x) = - f \rho \; \left\{ \left[ 1- s\, \tanh 
\rund{{x\over x_c}} \right] \, r^3 - r^{-(\nu+2)} \right\} \; , 
\label{e5.1}
\end{equation}
with $\rho$ as defined in this paper, and $\nu = 5-6$
or $\nu=-3$ for main sequence and giant companions respectively.
Some of the results given below depend on having $\rho>>1$, 
which is satisfied very well by giants 
[for which $\rho\approx\rho_{\rm nuc}$,
see (\ref{e3.2.4})] but not always by main--sequence companions.
Nevertheless, since most of the
analytic results described below can also be obtained for main--sequence 
companions, we shall quote them here too.

From (\ref{e2.11}) the critical curve $\dot x=0$ is given by
$1+p(r, x)-x=0$.
The chosen form of $p(r,x)$ gives two branches
of this curve with different slopes:
a low branch with $x<<x_c$ and a high branch where $x>>x_c$. 
For the low branch one can easily show
using the approximation $\tanh{(x/x_c)}\approx x/x_c$ that 
\begin{equation}
x_L \approx {f\rho(r^3-r^{-\nu-2}) -1
\over f\rho s r^3/x_c -1} \; .
\label{e5.2}
\end{equation}
The denominator in the above equation can be rewritten using (\ref{e4.2})
as $p_x \cosh^2{(x/x_c)}-1$. 
The condition $p_x>1$ for instability ensures that this 
denominator is positive near $x_0$;
as $x_L$ must be finite the denominator must remain positive everywhere
on the lower branch, implying a positive slope there.
Since in general the radial variations are small
departures from equilibrium, we can linearize around $r=1$ to get
$x_L=0$ at $r=1+1/[f\rho(5+\nu)]$. In general the slope of this
critical curve is given by 
\begin{equation}
\rund{\dd x\over \dd r}_{\dot x =0} = {-p_r\over p_x-1}\; ,
\label{e5.2a}
\end{equation}
where $p_r<0$ for physically relevant cases.
If $f\rho>>1$, as in the case of evolved companions, then
the slope of the lower branch at $x=0$ is approximately 
$(5+\nu)x_c/s$.

The upper branch is obtained by setting the
tanh factor to unity and thus 
\begin{equation}
x_U \approx 1+f\rho s r^3 -f\rho(r^3-r^{-\nu-2})\; . 
\label{e5.3}
\end{equation}
For
small departures from the equilibrium radius this reduces to
$x = 1+f\rho s -f\rho(5+\nu-3s)(r-1)$ showing that the upper branch has a 
large negative slope and is therefore stable (as $s < 0.5$ we have
$(5 +\nu-3s) > 3.5 + \nu$, which is $>0.5$ even for giant companions
with $\nu=-3$).
At some intermediate value of the transfer rate $x_t$ there is
a turning point at which $(\dd r/\dd x)_{\dot x=0} = 0$, or
equivalently where $p_x=1$. This turning
point therefore satisfies the equation 
\begin{equation} 
\cosh^2{\rund{{x_t\over x_c}}} = {sf\rho r_t^3\over x_c}\; ,
\label{e5.4}
\end{equation}
where $r_t$ is the dimensionless radius at the turning point. In
general $r_t$ and $x_t$ must be obtained by solving equation 
(\ref{e5.4}) together with $1+p(r,x)-x=0$, which can be easily done
iteratively. Note also that as in Paper I, the axis $x=0$ is formally
part of the $\dot x=0$ curve, although it takes infinite time to reduce
$x$ to zero (see Fig. 1). From the considerations above we
can see that the fixed point is unstable when $x_t>x_0>0$. 

The maximum expansion stage of the secondary is reached close to 
$r_t$. We could estimate $r_t$ by asking for the radius at which
$x_L=x_U$, but we prefer to use a physical
argument based on the assumptions made in the model:  
the maximum expansion of the companion is reached when all the
luminosity generated is emitted by the unilluminated side at the
unperturbed effective temperature. This yields $p(r_t,x_t)\approx 0$
and 
\begin{equation}
r_t \approx (1 - s_{\rm eff})^{-1/(\nu+5)} 
        \approx 1 + {s_{\rm eff}\over \nu + 5}\; ,
\label{e5.5}
\end{equation}
where $s_{\rm eff}=s \tanh{(x_t/x_c)}\sim s$. Formally if $p(r,x)=0$ 
for
non--vanishing illumination we again get the adiabatic mass transfer
rate $x=1$ and thus $\dot x=0$. This simple
argument shows that main sequence companions expand relatively little
under illumination while giants expand substantially:
in general we have $r_t \sim 
1 + 0.1s_{\rm eff}, 
1 + 0.5s_{\rm eff}$ for these two type of companion.
Thus with $s_{\rm eff}=0.3$ we find from (\ref{e5.5}) maximum radii 
$r_t({\rm ms}) = 1.036, r_t({\rm g}) = 1.195$ for main--sequence and 
(sub)giant companions respectively ($\nu = 5, -3$). Physically
this happens because upon expansion the nuclear luminosity of a 
main--sequence
star decreases, whereas the luminosity generated by a giant does not
depend on the radius. Thus assuming that the stars remain on the 
Hayashi line, a giant must expand until the unblocked
area equals its original surface area, while a main sequence star
finds a new irradiated equilibrium by expanding and simultaneously 
reducing its nuclear luminosity. Since $x_U = 1+p(x_U, r)$, 
the radius (\ref{e5.5}) ($p=0$) has $x_U=1$
with $s_{\rm eff}=s$,
while the radius at which $x_U$ and $x_L$ are equal differs 
from this only by 
terms $\sim [f\rho(\nu+5)]^{-1}$. We can then show from equation
(\ref{e5.4}) that $x_t$ is never very large, typically $\sim$ a 
few times the adiabatic rate.  

We can also make some general statements about the properties of the
other critical curve $\dot r=0$. 
All physically plausible $p(r,x)$
must be such that under no mass transfer and no illumination, the 
star remains at its equilibrium radius, implying $p(1,0)=0$.
Therefore from equation (\ref{e2.12}) one sees that the critical
curve $\dot r=0$ must go through the point $r=1, x=0$. The slope of
this curve at that point is also positive, but smaller than the slope
of the $\dot x=0$ curve at slightly larger $r$
if a fixed point with $x_0>0$ exists. This is
obvious from geometrical arguments, but can also be shown explicitly.

A typical limit cycle is shown in Fig. 1,
where we have stretched the $r$
variable and exaggerated the separation between the two critical
curves for the sake of clarity. Given that the $\dot r=0$ curve
intersects the $r$-axis at $r=1$, the entire cycle satisfies $r>1$,
i.e. the star is always somewhat oversized.
The critical curves cross at the
unstable fixed point $(r_0,x_0)$. The four points labelled ABCD along
the cycle identify the locations at which the phase point crosses 
critical 
curves. These naturally divide the cycle into four phases which we
discuss in turn. At point A the mass transfer reaches its
peak value $x_A(-\dot M_2)_{\rm ad}$ and we have maximum contact: 
$R_2-R_L\approx H \ln{(x_A/x_0)}$. We can estimate $x_A$ from the fact
that A lies almost vertically above the point where $x_L=0$. From
(\ref{e5.2}, \ref{e5.3}) we have $x_A\approx f\rho s$. In
physical units this rate is 
\begin{equation}
(-\dot M_2)_{\rm max}\approx x_A {M_2\over (\zeta_s -\zR)t_{\rm dr}} =
{s M_2\over (\zeta_s -\zR)t_{\rm ce}} \; ,
\label{e5.7}
\end{equation}
where $t_{\rm ce}$ is the timescale for thermal relaxation of the
convective envelope, which can be significantly shorter than $\tKH$.
At point B the companion is very
close to its maximum size, while the binary
orbit has expanded so that $R_2\approx R_L\approx r_t R_e$. The
degree of overfilling of the Roche lobe has been reduced to 
$R_2-R_L\approx H\ln{(x_t/x_0)}$. The stellar expansion rate 
$\dot R_2/R_2=t_{\rm nuc}^{-1}$ is now too slow 
compared with that of the lobe to sustain the high mass transfer
rate which in turn drives the radius expansion. 
(This is inevitable since 
the transfer rate cannot indefinitely remain above the secular 
mean driven by nuclear expansion and angular momentum losses.) 
Thus the companion rapidly loses
contact and the mass transfer drops very sharply,
while the star shrinks back towards
its equilibrium radius. At point C minimum mass transfer is reached
because the system is maximally detached, with 
$R_2-R_L\approx -(r_t-1)/\epsilon$, so that $x$ is essentially zero in
the low state. At this point $\dot R_2$ is again very close to zero,
and nothing will happen until combined effects of
nuclear evolution and angular momentum losses bring the system
close to contact again. At D the secondary is expanding slightly 
($\dot R_2/R_2 = \dot R_e/R_e$) so that $\dot r=0$. 
The companion radius $R_2$ is now within a few
scale heights of 
$R_L$. This raises the transfer rate, making the companion
expand more rapidly under
irradiation, which in turn increases the transfer rate, so that the 
cycle restarts. 

We can also estimate the timescales for
the different phases of the cycle and evaluate the total mass
transferred in a cycle. We estimate the rise time by
assuming that the mass transfer initially increases because the
thermal imbalance due to irradiation forces the secondary into deeper
contact. The characteristic rate of radial expansion when a
fraction $s$ is fully blocked is 
$\dd\ln{R_2}/\dd t\approx s/t_{\rm ce}$.
Therefore the rise time is approximately
the time required to expand by $\ln{(x_A/x_0)}$ 
scale heights
\begin{equation}
t_{DA}\approx {\ln{(x_A/x_0)} H\over R_2}{t_{\rm ce}\over s}
\qquad {\rm or} \qquad \tau_{DA} \approx {\ln{(x_A/x_0)}\over x_A}\; ,
\label{e5.8}
\end{equation}
where we have used our estimate of $x_A=sf\rho$ to obtain the final
expression. 
During the time $t_{AB}$ spent on the high branch, the rate of
expansion
of the secondary decreases monotonically from the maximum rate 
estimated above, which yields the peak mass transfer rate $x_A$,
until  $\dd\ln{R_2}/\dd t = 1/t_{\rm nuc}$. At that point (B) we have
$\dot r=0$ 
and the expansion rate falls below the driving rate so that the system
detaches, going rapidly into a low state. We adopt the following
ansatz for the radius as a function of time
\begin{equation}
R_2(t) = R_A + (R_B-R_A)(1-e^{-t/T_+})\; .
\label{e5.9}
\end{equation}
Equating  $\dd\ln{R_2(0)}/\dd t = s/t_{\rm ce}$ and  
$\dd\ln{R_2(t_{AB})}/\dd t = 1/t_{\rm nuc}$, we obtain both the
characteristic radial expansion timescale $T_+$ and the time 
$t_{AB}$ spent on the high branch
\begin{equation}
T_+ \approx {(r_t-1)\over s} t_{\rm ce} \sim {t_{\rm ce}\over \nu +5}
\qquad {\rm or} \qquad  
\tau_+ \approx {r_t-1\over\epsilon x_A}
\label{e5.10}
\end{equation}
and
\begin{equation}
t_{AB} \approx {(r_t-1)\over s} t_{\rm ce} \ln{(x_A/r_t)} 
\sim {t_{\rm ce}\over \nu + 5}\ln{(x_A/r_t)}
\qquad {\rm or} \qquad 
\tau_{AB} \approx {r_t-1\over\epsilon x_A} \ln{(x_A/r_t)} \; ,
\label{e5.11}
\end{equation}
where we have taken $R_A\approx R_e$ and $R_B\approx r_t R_e$.
As soon as the system detaches, irradiation ceases, and the companion
contracts rapidly while the orbit and Roche lobe remain at the
size attained at the end of the high state. The contraction is even 
more rapid than the expansion because the star is more luminous. One
can show that the characteristic thermal contraction time scale is 
\begin{equation}
\tau_- \approx (1-s)^{3/(\nu+5)} \tau_+ \; ,
\label{e5.12}
\end{equation}
leading to 
\begin{equation}
t_{BC} \approx{r_t-1\over r_t^3}{t_{\rm ce}\over s_{\rm eff}}
\ln{r_t^2s_{\rm eff}t_{\rm dr}\over t_{\rm ce}} 
\qquad{\rm or} \qquad
\tau_{BC}\approx 
{(r_t-1)t_{\rm ce}\over r_t^3\epsilon s_{\rm eff}t_{\rm dr}}
\ln{r_t^2s_{\rm eff}t_{\rm dr}\over t_{\rm ce}}\; . 
\label{e5.12a}
\end{equation}
Clearly this effect is more pronounced in giant companions than in 
main--sequence secondaries, for reasons that have already been
mentioned. In a few times 
$\tau_-$ (i.e. typically a time $\lta 0.5t_{\rm ce}$)
the mass transfer formally 
reaches a minimum value $x_C\approx 0$ and 
a long detached (low) state now follows while the driving tries to
bring
the system back to contact. The time spent in the low state is thus 
dominated by the time $t_{CD}$ 
\begin{equation}
t_{CD} \approx (r_t-1)t_{\rm dr} \sim {s\over \nu + 5}t_{\rm dr}
\qquad {\rm or} \qquad
\tau_{CD} \approx {(r_t-1)\over \epsilon} \; .
\label{e5.13}
\end{equation}

The total mass transferred during a cycle can now be estimated as
$\Delta M_2 \approx x_A (-\dot M_2)_{\rm ad} T_+ $, which yields 
the simple -- with hindsight perhaps obvious -- result
\begin{equation}
\Delta M_2 \approx (-\dot M_2)_{\rm ad} (r_t-1) t_{\rm dr} 
     = {r_t-1\over \zeta_s - \zeta_R}M_2 \approx {s_{\rm eff}\over 
(\zeta_s - \zeta_R)(\nu+5)}M_2\; .
\label{e5.14}
\end{equation}
Hence the total cycle time $t_{\rm cycl} \simeq 
\Delta M_2 / (-\dot M_2)_0$ is 
\begin{equation}
t_{\rm cycl} \simeq \frac{r_t -1}{x_0} \: t_{\rm dr} 
\qquad {\rm or} \qquad
\tau_{\rm cycl} \approx {(r_t-1)\over \epsilon x_0} \; .
\label{e5.15}
\end{equation}
Thus irradiated main--sequence
stars ($\nu = 5 - 6$)
transfer at most a few percent of their mass per cycle. 
In contrast giants ($\nu = -3$) transfer a significant 
fraction $\sim s$ of their total mass $M_2$ in the high state, which
may amount to most of the convective envelope. Since
$t_{\rm ce}$ is also much shorter for giants we expect
that if the irradiation instability is allowed to grow unchecked in 
such systems it
will produce accretion rates which are super--Eddington in LMXBs. The
resulting common--envelope evolution would probably lead to the 
formation of an ultrashort--period binary.
Clearly the irradiation instability must be quenched in observed LMXBs 
with evolved companions. In Section 7 we 
discuss ways in which this can happen. For reference we collect together
the analytic expressions for the various properties of the limit cycle
in Table 1.

\section{NUMERICAL RESULTS}

We have integrated the evolution equations (\ref{e2.11}, 
\ref{e2.12}) from
arbitrary initial states for a variety of parameter choices. In this 
section
we present a few examples and compare them with the analytic
estimates given in the previous section. The values quoted 
in the text are those obtained by numerical integration, with
the corresponding analytic estimate in brackets. The 
analytic estimates for peak rates,
rise times, and maximum expansion radius are in all cases calculated 
in very good agreement with numerical results. In general we 
integrated the equations for several cycles (3-6) and noted that the
first outburst is usually somewhat untypical (slightly higher for the
same radii and a higher peak rate) because of initial conditions,
whereas later outbursts are virtually identical. 
We therefore quote values taken from later outbursts.

Fig. 2 shows a case with core mass $M_c = 0.15\Msun$, the
lowest reasonable value for which the approximations used in Section 3
are still valid. We have also taken $s=0.5$, which is perhaps
unrealistically large unless there is significant scattering from a
disc corona or similar structure. The luminosity and radius of 
the companion are at the
lower end of the subgiant range and therefore the instability is
weakest. Nevertheless a very large amplitude outburst results, with a
peak transfer rate $x_A = 152.8$ (154.9) which is attained rapidly, in
$\tau_{DA} = 0.032$ (0.033). The companion continues to expand in the
high state until the turning point is reached at $r_t= 1.41421$
(1.41421). 
The duration of the high state can be read off the graph directly and 
is $\tau_{AB}= 8.1$ (12.5). While the behaviour of $r$ as shown in
Fig. 2 appears to obey equation (\ref{e5.9}), closer examination
shows that the expansion phase is actually faster than exponential
while the contraction phase is slower than the assumed exponential.
Nevertheless we have estimated  the characteristic radial expansion
and contraction timescales graphically obtaining $\tau_+ = 2.4$ (2.67)
and $\tau_- = 1.0$ (0.95). Despite the fact that equation (\ref{e5.9})
holds only approximately, the expressions based on it are better than
order of magnitude estimates. 
For the case shown, the total duration of the cycle is 415.9 and 
the duration of the low state is $\tau_{CD}\approx 407.8$ (414.2). 

Figure 3 shows another case with a higher core mass $M_c=0.25 \Msun$ 
and $s=0.3$, appropriate for illumination by a point source. 
The instability in this case is more violent, with shorter rise times 
$\tau_{DA}= 0.0023$ (0.0024)
and higher peak rates $x_A = 3326$ (3609), but a smaller turning radius
$r_t=1.19529$ (1.19523). The total duration $\tau_{AB}=0.45$ (0.43)
of the high state is
relatively short. The characteristic e-folding
times for radial expansion $\tau_+ = 0.051$ (0.054) and contraction
$\tau_- = 0.04$ (0.031) are also relatively shorter because of the
higher luminosity.
The analytic estimates for the properties of the radial variations
during the limit cycle are even more accurate than in the case in Fig.2,
because the 
neglected terms $\sim (f\rho s)^{-1}$ are still smaller here.
The total duration of the cycle is 188.6 and the low state lasts for 
$\tau_{CD}\approx 188.1$ (195.2).

\section{THE IRRADIATION INSTABILITY IN LOW MASS BINARIES}

We can now discuss the application of the theory developed here to various
types of close binary encountered in nature. The only restriction
is that the companions must have a significant convective
envelope and thus a low mass $M_2 \lta 1.5\msun$.
The instability criterion implying cycles given in 
(\ref{e4.3}) can be rewritten as 
\begin{equation}
f\rho s > {x_0\over r_0^3} {x_c\over x_0} 
\cosh^2{\biggl({x_0\over x_c}\biggr)} \; .
\label{e7.1}
\end{equation}
Here the factor $x_0/r_0^3\sim 1$ does not vary much, whereas 
$f\rho s$ is mainly sensitive to the type of companion and the mechanism 
driving the binary evolution, while $\xi=x_0/x_c$ depends on both the 
accretor and donor type. 
In Fig. 4 we plot $x_0/x_c$ along the abscissa (the compact object
axis) and $f\rho s$ along the ordinate (the companion axis).  
Changing the type of companion causes displacements along both axes,
whereas changing the primary causes only horizontal displacements.
The stability/instability boundary plotted is simply
$f\rho s = \xi^{-1}\cosh^2{\xi}$. The locations of various possible
evolutionary sequences for CVs and LMXBs are also shown on
Fig. 4. Changing $s$ by screening or scattering causes
purely vertical displacements since $\xi$ is not affected: clearly for
any binary there is a value of $s$ below which mass transfer is stable.
Within the limitations of the simple theory developed here
(i.e. the form chosen for $p(r, x)$) the results displayed on Fig.4 show 
the following:

\noindent
1. CVs above the period gap and above a certain period (or companion
mass) can be unstable, depending on the value of $\eta$ 
(see Paper I). CVs below the gap are stable.

\noindent
2. Long period CVs with companions having 
small core masses are stable, whereas companions with larger core masses
likely to have even longer orbital periods are unstable. 
A detailed analysis shows that
GK Per and V1017 Sgr are unstable if $\eta\gta 0.08$ and 0.04 respectively.

\noindent
3. Short--period 
LMXBs with main sequence or partially evolved
companions are stable because they 
are so strongly irradiated that they have reached saturation
($x_0 >> x_c$).
Thus variations in $x$ do not cause radius variations, eliminating the
feedback necessary for instability.

\noindent
4. Long period LMXBs with (sub)giant companions are unstable:
the larger the core mass and the smaller the total companion mass, 
the more violent the instability. 

The irradiation instability 
may well cause the formation of ultrashort--period systems 
(e.g. the AM CVn's) from long--period CVs and LMXBs. However,
since the rise to the high state is so rapid it is extremely improbable
that we currently observe any long--period LMXB undergoing cycles. We 
must therefore consider ways of quenching the instability in these
systems. The most obvious possibility is screening of the companion
from the irradiation, which is the basic cause of the instability. 
Screening may well result from the extensive disc coronae inferred in
LMXBs (e.g. White \& Mason, 1985). Formally we can consider this
possibility by decreasing $s$ in our simulations.
As can be seen from equation (\ref{e5.4}) the effect of a small $s$
is to lower the high branch and thus reduce the amplitude of the
cycle in $x$. Since the lower branch remains unaffected, the
turning radius and hence the radial amplitude will also be reduced.
This suggests that screening could reduce the amplitude of the
cycle and increase the frequency of outbursts, because it will take
less time to drive the system into contact again once it has detached
following an outburst. However this argument applies only to a fixed $s$.
If $s(x)$ is itself a
function of the instantaneous mass transfer rate, as a result of a
varying geometrical thickness of the accretion disk or varying
optical depth through the (out)flow, a more gradual transition
occurs. For example, if screening results in a rapid reduction of $s$
beyond some critical $x_{\rm scr}$, the amplitude of the cycle in $x$ 
is reduced, saturating somewhere close to $x_{\rm scr}$, whereas
the radial amplitude remains large; thus the outburst does not recur
any sooner than in the unscreened case. We have performed some
numerical experiments to simulate these effects and verify the above
statements. These simulations and the discussion above 
imply that if  $x_{\rm scr}>x_t$, the radial amplitude of the cycle and
recurrence time are not significantly different from the unscreened
case. However, if $x_{\rm scr}<x_t$, then both the radial amplitude
and the duration of the low state are reduced. Also the duration of
the high state and the amount of mass transferred per cycle are 
decreased. Clearly if $x_{\rm scr}$ is reduced to $\sim x_0$ we will find
that the cycles disappear altogether. Thus efficient screening can either
eliminate the cycles entirely or render them relatively harmless as far as
the binary evolution is concerned.

While screening must play a role in stabilizing some systems against the
irradiation instability, a second way of quenching the instability 
appears
to be more common. This mechanism uses the fact that the companion
swells only when its own intrinsic luminosity is blocked by the 
irradiating flux, and that the blocking effect saturates once the latter
is comparable with the intrinsic flux. A variable accretion rate, as 
seen in e.g. soft X--ray transients, severely reduces the efficiency of
irradiation in expanding the companion: the very high irradiating flux
during outbursts has no more effect than a much weaker value, while the
star can cool between outbursts. We might thus expect a highly modulated
accretion rate with a duty cycle $d<<1$ to mimic irradiation by a steady
accretion rate a factor $d$ smaller. This expectation is largely 
fulfilled, as the following simple calculation shows.

We assume the dimensionless accretion rate to vary periodically in 
time over $t_{\rm rec}$ as
\begin{equation}
x_{\rm acc} = x_h, \qquad 0<t \leq t_h, \qquad = x_l, \qquad
t_h < t \leq t_{\rm rec}.
\label{e7.2}
\end{equation}
Mass conservation requires that the dimensionless transfer rate obeys
\begin{equation}
x= dx_h+(1-d)x_l=x_h\biggl[d+ {1-d\over A}\biggr] \label{e7.3}
\end{equation}
where $d=t_h/t_{\rm rec}$ is the duty cycle and $A=x_h/x_l$ the 
amplitude of the accretion rate variation. The reaction of the
companion to this intermittent irradiation is governed by the 
thermal relaxation function (\ref{e5.1}), which now becomes
\begin{equation}
p=-f\rho\; \left\{\left[1-s\, \tanh\rund{{x\over x_c}{A\over d(A-1)+1}}
\right]\, r^3-r^{-(\nu+2)}\right\}, \qquad 0<t\leq t_h\; ,
\end{equation}
\label{e7.4}
and
\begin{equation}
p=-f\rho\; \left\{\left[1-s\, \tanh\rund{{x\over x_c}{1\over d(A-1)+1}}
\right]\, r^3-r^{-(\nu+2)}\right\}, \qquad t_h<t\leq t_{\rm rec}\; .
\label{e7.5}
\end{equation}
Now assuming that the cycle time $t_{\rm rec}$ is $<<t_{\rm ce}$, we can
define a mean value 
\begin{equation}
<p>={1\over t_{\rm rec}}\int_0^{t_{\rm rec}}p(x, r){\rm d}t\; .
\label{e7.6}
\end{equation}
Performing the integration, we can compute the derivative $<p>_x$ which
governs the stability of the fixed point $(x_0, r_0)$.
If $x_h >> x_l$, i.e. $A\rightarrow \infty$, as is characteristic
of soft X--ray transients and dwarf novae, we have
\begin{equation}
<p(x_0, r_0)>_x ={f\rho s r_0^3\over x_c}{\rm sech}^2\rund{{x_0\over dx_c}}\; ,
\label{e7.8}
\end{equation}
and the criterion for instability $<p(x_0, r_0>_x > 1$ can be written as
\begin{equation}
f\rho s > {x_0\over r^3}\rund{{x_c\over x_0}}{\rm cosh}^2\rund{{x_0\over 
dx_c}}\; .
\label{e7.9}
\end{equation}
This criterion is very similar to the earlier one (\ref{e7.1}) 
assuming steady accretion, to which it of course reduces as 
$d \rightarrow 1$. For $x_0/dx_c \lta 1$ the two criteria are 
identical. Thus on Fig. 4 the stability/instability boundary for $d < 1$
is given simply by sliding the $d=1$ curve parallel to itself along
the asymptote for small $x_0/x_c$, by a displacement of $-\log d$ in 
$x_0/x_c$. We may thus draw a further conclusion from Figure 4: 

\noindent
5. Typical soft X--ray transient
duty cycles $d \lta 10^{-2}$ are probably enough to stabilize most
LMXBs with evolved companions against the irradiation instability. 

By contrast, extremely short duty cycles $d \lta 10^{-4}$ would be required
for dwarf nova outbursts to stabilize CVs with evolved secondaries. We
note that most LMXBs with periods $\gta 1$ d are transient (King, Kolb
\& Burderi, 1996), and indeed this holds for a very large fraction
of long--period systems also (King, Kolb \& Burderi, 1996; 
King et al., 1996b). We shall consider the application of the stability
criteria to individual systems in a future paper.

\section{CONCLUSIONS}

We have shown that irradiation of an evolved low--mass companion in LMXBs
and CVs can drive mass transfer cycles. These cycles do not need the
intervention of any further effect, unlike the case of main--sequence
companions (Paper I), where a modest increase in the driving angular 
momentum loss rate is required in the high state if cycles are to occur
in many cases.
The cycles with evolved companions are also considerably more violent than
in the main--sequence case. This is a direct consequence of  
two facts. First,
the nuclear luminosity of an evolved star is insensitive to the stellar
radius, so that blocking of the intrinsic stellar flux by irradiation
requires the star to expand so as to maintain the same unblocked area. This
leads to much larger expansions than in the main--sequence case, where the
nuclear luminosity drops sharply as the star expands. 
Second, the ratio $t_{\rm dr}/t_{\rm ce} = f\rho$ of driving to thermal 
timescales 
is much larger in evolved stars than on the main sequence, making the
expansion very rapid. In the high state an evolved star 
would lose a significant fraction of its total envelope mass on a thermal
timescale. 

In a CV with a low core mass, 
the implied accretion rate would probably turn the 
system into a supersoft X--ray
source. In CVs with higher core masses, the nuclear burning causes the
white dwarf to develop an extensive envelope, while in long--period
LMXBs, the high state
accretion rates greatly exceed the Eddington limit. If the
instability is not quenched all 
these systems would undergo a common--envelope
phase. They may merge entirely, or reappear
as ultrashort--period systems like AM CVn
(for white dwarf primaries), or helium--star LMXBs, or detached
systems with low--mass white dwarf companions.
However, there
are at least two ways in which the instability can be quenched and
the systems (particularly LMXBs)
stabilized: shielding of the companion by e.g. an
extensive accretion disc corona, and intermittent accretion. The typical
duty cycles $d \lta 10^{-2}$ observed in soft X--ray transients are 
short enough to stabilize them, while observed dwarf nova duty cycles 
are unable to stabilize CVs with evolved companions.  

Although both means of stabilization seem to occur in nature, it is
clear that the irradiation instability is so violent that it must play
a major role in any discussion of the evolution of CVs and LMXBs with
evolved companions: the systems must somehow stave it off or evolve
catastrophically. We shall consider some of the observational 
consequences in a future paper.

JF thanks the Max--Planck--Institut f\"ur Astrophysik for warm
hospitality during a productive stay in June--August 1996. This
work was partially supported by the U.K. Science and Engineering
Research Council (now PPARC) and by NASA grant NAG5-2777 to LSU. 
ARK acknowledges support as a PPARC Senior Fellow, and the warm hospitality
of the MPA.

{}

\clearpage

\renewcommand{\arraystretch}{1.5}

\begin{table}
\caption{\bf Characteristic Properties of the Limit Cycle}
\bigskip
	
\begin{tabular} {ll}
{\bf Parameters} & \\
\tablevspace{5pt}\tableline\tablevspace{3pt}
$\tdr = (1/t_{\rm nuc} + 2/t_J)^{-1}$ 		& driving timescale  \\
$\tce = f^{-1}\tKH$                			& thermal timescale of 
the convective envelope \\
$\zeta_s = (\partial\ln{R_2}/\partial\ln{M_2})_s$	& adiabatic mass radius 
exponent \\
$\zeta_e = (\partial\ln{R_2}/\partial\ln{M_2})_e$	& thermal equilibrium 
mass radius exponent \\
$\zeta_R = (\partial\ln\RL/\partial\ln{M_2})$		& Roche lobe mass radius 
exponent \\
$\epsilon = H/R_{2,e}$					& photospheric
scale height in units of equlibrium radius \\
$\nu = \left\{ \begin{array}{rl}  5-6 & \mbox{MS donors} \\
 -3 & \mbox{giant donors} \end{array} \right.$
&  $(-\nu-3)$ is the radius exponent of the nuclear luminosity\\
$\se = s\:\tanh (x_t/x_0) \simeq s \simeq 0.3$ & effective blocked
surface fraction \\
\tablevspace{3pt}\tableline\tablevspace{10pt}
{\bf Characteristic Properties} in the limit & $\se \tdr/\tce \gg 1$ \\
\tablevspace{5pt}
in physical units  & in dimensionless form \\
\tablevspace{5pt}
\tableline
\tablevspace{5pt}
$(-\dot M_2)_0 = \DS\frac{\DS M_2}{\DS \zeta_e-\zR}\:\DS\frac{\DS 1}{\tdr}$ 
& $x_0 = \DS\frac{\dzsR}{\DS \zeta_e-\zR}$ \\
\tablevspace{5pt}
$(-\dot M_2)_A \approx\DS\frac{\DS M_2}{\dzsR}\:\DS\frac{\se}{\tce}$ 
& $x_A \approx\DS\frac{\se\tdr}{\tce}$ \\
\tablevspace{5pt}
$R_{2,t} \equiv R_{2,B}\approx (1-\se)^{-1/(\nu+5)} R_e$ 
& $r_t \equiv r_B \approx (1-\se)^{-1/(\nu+5)}$ \\
\tablevspace{5pt}
$\Delta M_2\approx \DS\frac{\rt1}{\dzsR} M_2$ & $\DS\frac{\DS \Delta 
M_2}{\DS M_2}\approx\DS\frac{\rt1}{\dzsR}$ \\
\tablevspace{5pt}
$t_{AB}\approx\DS\frac{\rt1}{\se}\:\tce\:\ln\DS\frac{\se\tdr}{\DS r_t\tce}$	
& $\tau_{AB}\approx\DS\frac{\DS \rt1}{\DS 
\epsilon\: x_A}\:\ln\DS\frac{\DS x_A}{\DS r_t}$ \\
\tablevspace{5pt}					 
$t_{BC}\approx\DS\frac{\rt1}{\DS r_t^3}\:\DS\frac{\tce}{\se}\:\ln\DS\frac{\DS 
r_t^2\se\tdr}{\tce}$	& $\tau_{BC}\approx\DS\frac{\DS 
\rt1}{\DS r_t^3\epsilon\:x_A}\:\ln\DS r_t^2 x_A$ \\
\tablevspace{5pt}					 
$t_{CD}\approx (\rt1)\:\tdr$			& 
$\tau_{CD}\approx\DS\frac{\rt1}{\DS \epsilon}$ \\
\tablevspace{5pt}					 
$t_{DA}\approx\DS\frac{\DS \epsilon\tce}{\se}\:\ln\DS\frac{\se\tdr}{\DS 
x_0\tce}$	&
$\tau_{DA}\approx\DS\frac{\DS 1}{\DS x_A}\:\ln\DS\frac{\DS x_A}{\DS x_0}$ \\
\tablevspace{5pt}
$t_{\rm cycl}\approx\DS\frac{\rt1}{\DS x_0} \: \tdr$	
& $\tau_{\rm cycl}\approx\DS\frac{\rt1}{\DS \epsilon x_0}$ \\
\tablevspace{5pt} 
\tableline
\end{tabular}

\end{table}

\clearpage
\begin{figure}
\vspace*{-2.5cm}
\plotone{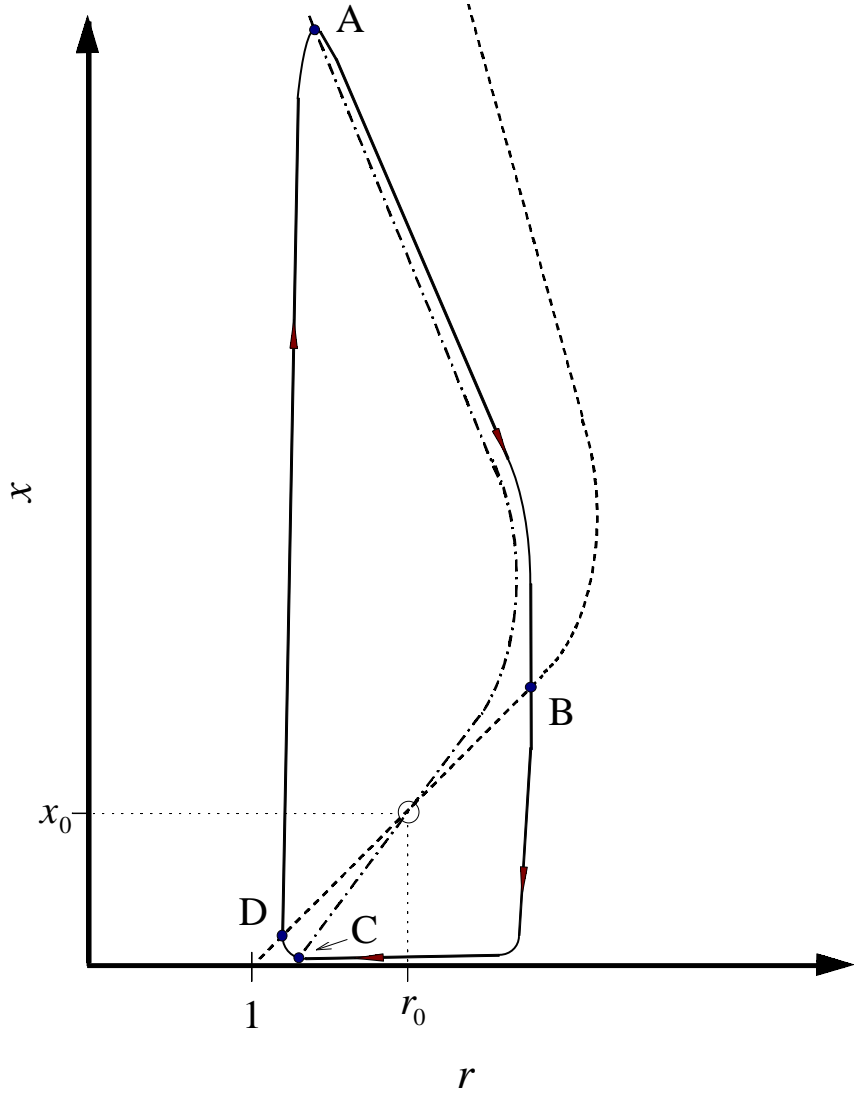}
\vspace{-1cm}
\caption{ {\small
Phase plane for the ODE system (\ref{e2.11}, \ref{e2.12}). 
The two critical curves where $\dot x = 0$ (dot-dot-dashed)
and $\dot r =0$ (dashed) are shown
for a typical $p(r,x)$. These curves intersect at
the fixed point $(r_0,x_0)$ and divide the $(r, x)$ plane into four regions
in which the motion of the system point is indicated by the arrows. 
The limit cycle intersects the critical curves at the points ABCD
giving the four phases of the cycle discussed in the text: a high
state AB during which the companion expands, a moderately rapid
contraction phase BC, a long low state CD, and an extremely fast rise DA
to peak mass transfer.}
}
\end{figure}

\begin{figure}
\vspace*{-2.5cm}
\plotone{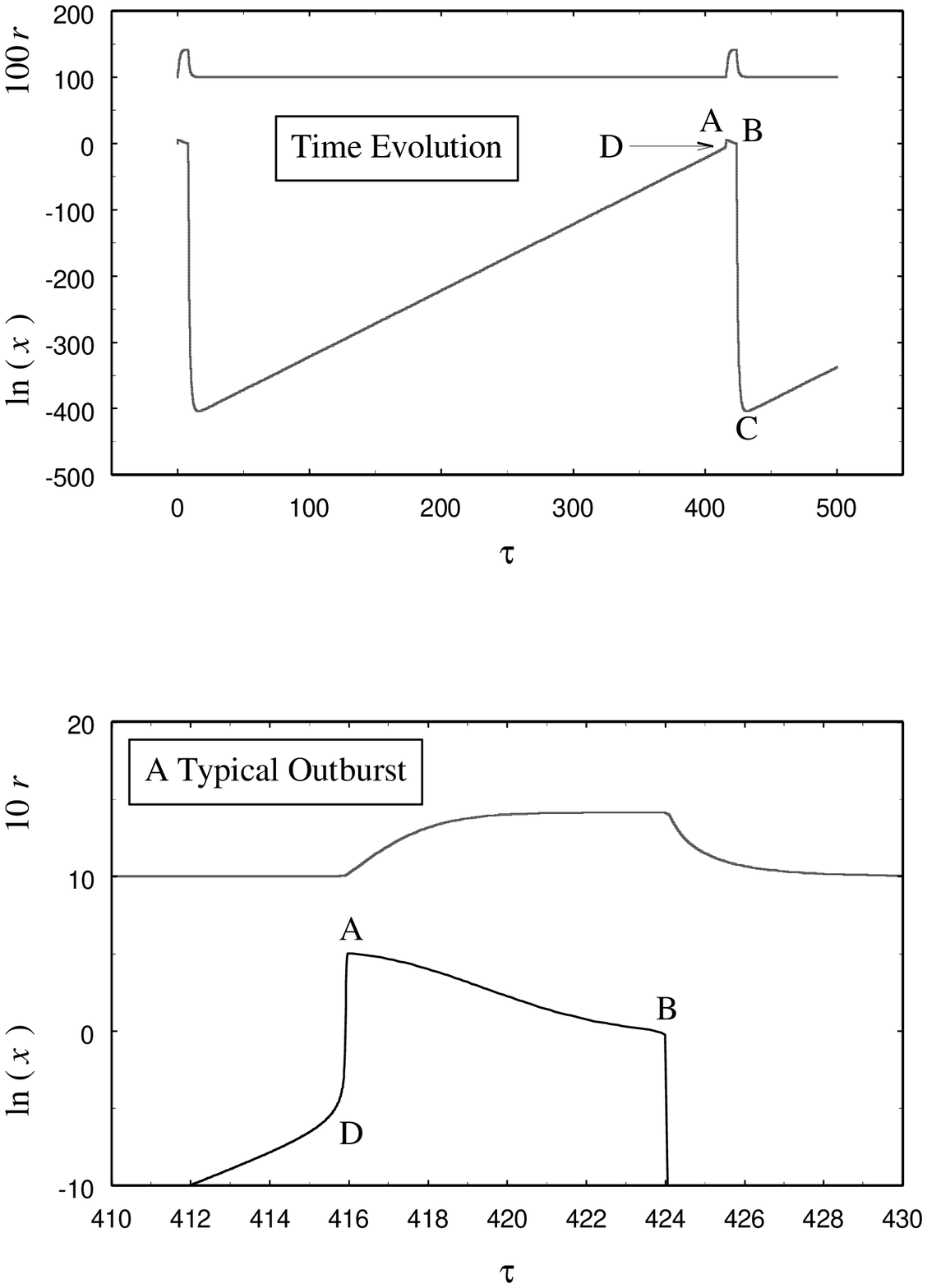}
\caption{ {\small
The dimensionless mass transfer $x$ and radius $r$ evolution during
a typical outburst for a subgiant companion with core mass 
$M_c= 0.15 \Msun$ and a large illuminated fraction $s=0.5$. The
calculations assumed a fixed $t_J=10^{10}$ yr, a neutron star primary
with $M_1 = 1.4 \Msun$ and $R_1=10^6$ cm, $M_2=0.5\Msun$, 
$\epsilon = 0.001$, $k\eta = 0.01$, $\zeta_s = -1/3$, $\zeta_e = -0.2$,
and $\zR = 2M_2/M_1 -5/3$.  
}}
\end{figure}

\begin{figure}
\vspace*{-2.5cm}
\plotone{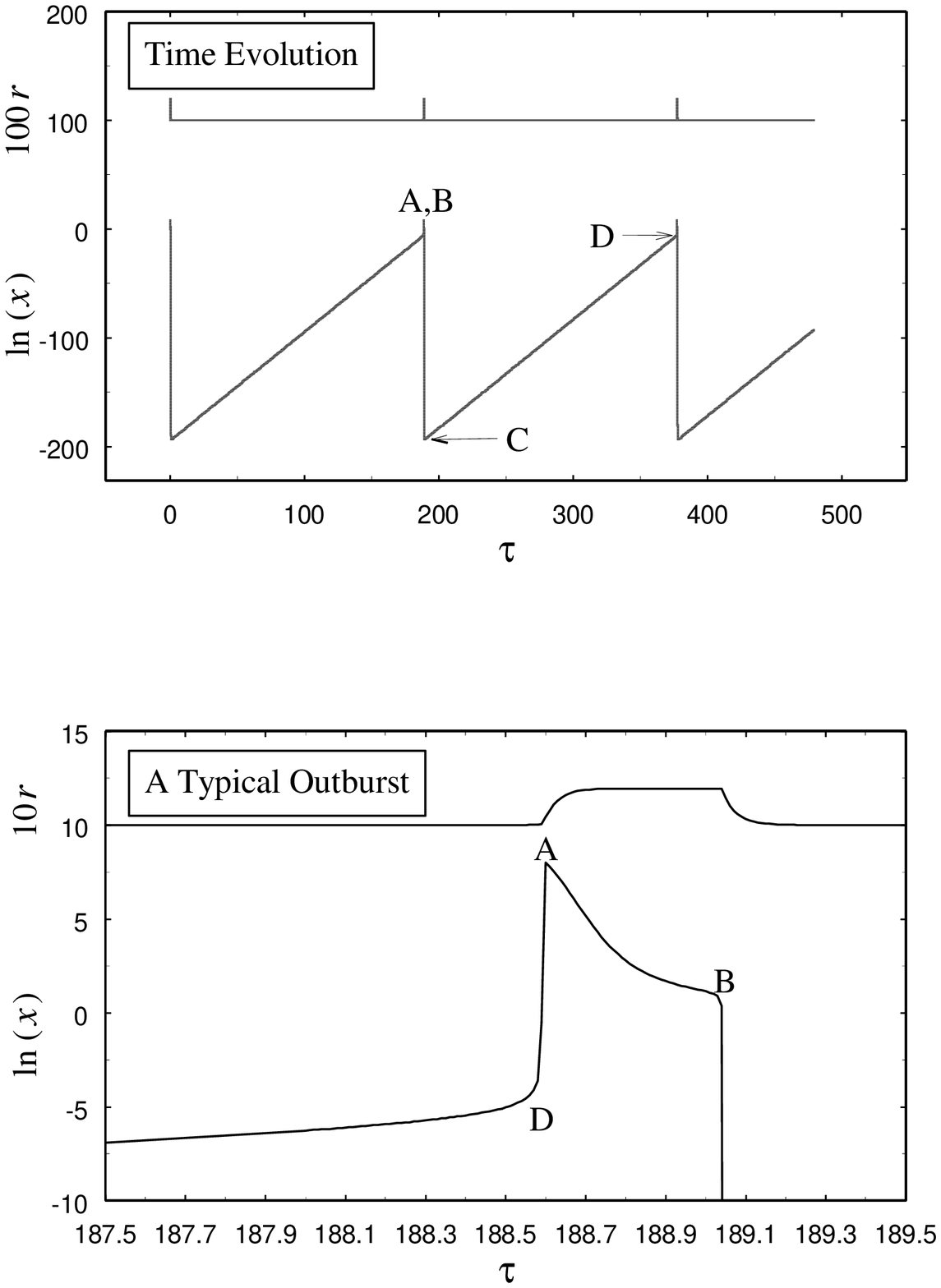}
\caption{ {\small
The dimensionless mass transfer $x$ and radius $r$ evolution during
a typical outburst for a giant companion with core mass 
$M_c= 0.25 \Msun$ and an illuminated fraction $s=0.3$, consistent with
a point source. The
calculations assumed a fixed $t_J=10^{10}$ yr, a neutron star primary
with $M_1 = 1.4 \Msun$ and $R_1=10^6$ cm, $M_2=0.5\Msun$, 
$\epsilon = 0.001$, $k\eta = 0.01$, $\zeta_s = -1/3$, $\zeta_e = -0.2$,
and $\zR = 2M_2/M_1 -5/3$.  
}}
\end{figure}

\begin{figure}
\vspace*{-2.5cm}
\plotone{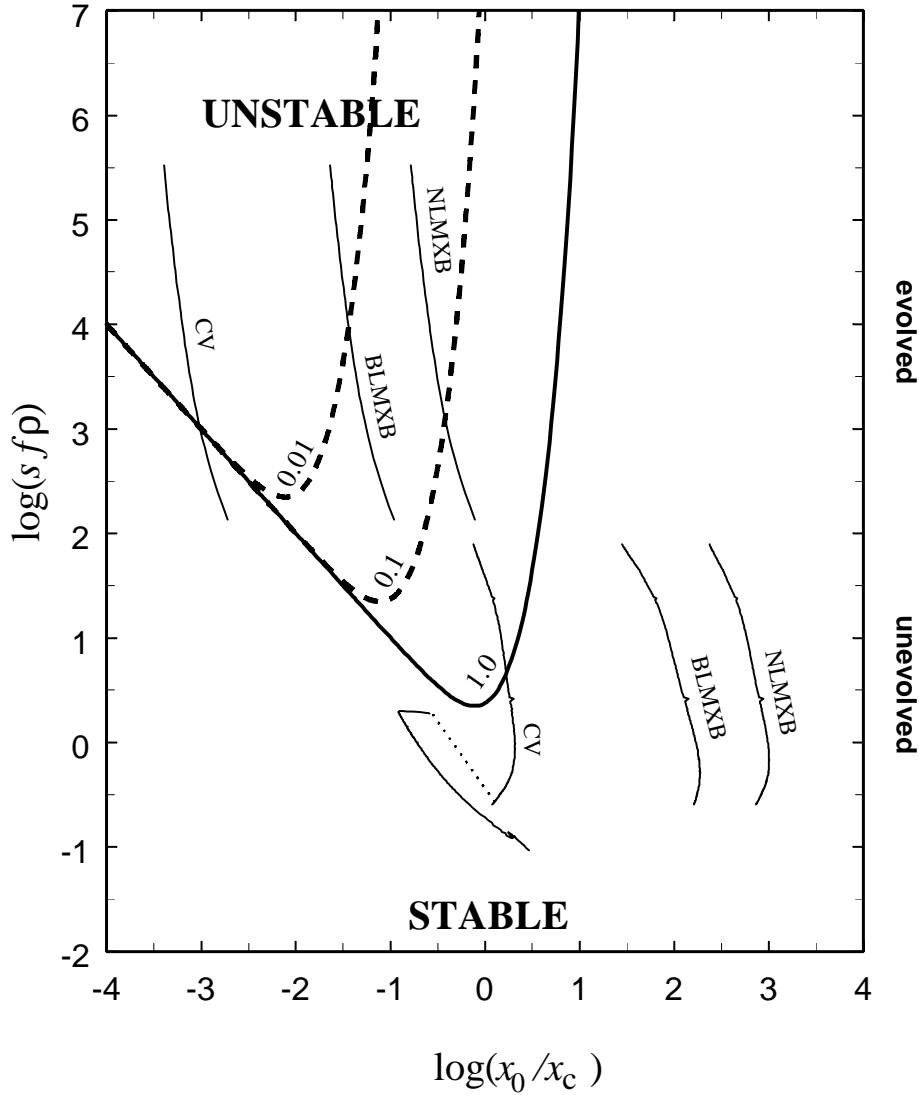}
\vspace{-1cm}
\caption{ {\small
The approximate stability boundary  given by equation (\ref{e7.9})
in the $(x_0/x_c,f\rho s)$-plane, for various values of the
accretion duty cycle $d$. Representative
evolutionary sequences for different 
types of low--mass binaries are also indicated. These include black holes,
neutron stars and white dwarfs 
with main--sequence and low--mass giant companions (denoted BLMXB, NLMXB,
CV respectively). 
We took black--hole, neutron star and white dwarf
masses $M_1 = 10\msun, 1.4\msun$ and $1.0\msun$, 
an irradiated fractional area $s = 0.3$,
$k\eta =0.1$ in short--period systems,
and $k\eta=0.01$ in long--period systems. The latter were evolved from
an initial core mass $M_{ci}= 0.15 \msun$ and total 
secondary mass $M_{2i} = 0.5 \msun$.
We see that steady LMXBs with evolved companions are 
prone to the irradiation instability, while CVs
with giant companions are unstable if their core masses are
sufficiently large (see text). Typical soft X--ray transient duty cycles
$d \lta 10^{-2}$ quench the irradiation instability in LMXBs,
while typical dwarf nova duty cycles cannot do this for CVs. We stress that
while the sequences shown here are representative, the stability or otherwise
of any given individual system must be considered in detail. 
}}
\end{figure}

\end{document}